\documentclass[10pt, conference, compsocconf]{IEEEtran}
\ifCLASSINFOpdf
   \usepackage[pdftex]{graphicx}
   \graphicspath{{figs/}}
   \DeclareGraphicsExtensions{.pdf,.jpeg,.png}
\else
\fi
\usepackage[cmex10]{amsmath}
\usepackage[caption=false,font=footnotesize]{subfig}
\usepackage{stfloats}

\usepackage{url}

\usepackage[draft]{changes}
\definechangesauthor[color=red]{rs}
\definechangesauthor[color=blue]{ya}

\usepackage{braket} 
\usepackage{bm} 
\newcommand{\vect}[1]{\boldsymbol{\mathbf{#1}}}

\newcommand{\HsubM}{\hat{H}_{\!M}}
\newcommand{\HsubC}{\hat{H}_{\!C}}

\newcommand{\Hc}{\hat{H}_{C}}
\newcommand{\Hm}{\hat{H}_{M}}
\newcommand{\D}{\mathcal{D}}
\newcommand{\norm}[1]{\left\lVert#1\right\rVert}

\usepackage{printlen} 

\makeatletter
\newcommand\thefontsize[1]{{#1 The current font size is: \f@size pt\par}}
\makeatother

\begin{document}
\bstctlcite{IEEEexample:BSTcontrol} 

%
\title{Evaluating Quantum Approximate Optimization Algorithm: A Case Study}


\author{\IEEEauthorblockN{Ruslan Shaydulin}
\IEEEauthorblockA{School of Computing\\
Clemson University\\
Clemson, USA\\
Email: rshaydu@g.clemson.edu}
\and
\IEEEauthorblockN{Yuri Alexeev}
\IEEEauthorblockA{Computational Science Division\\
Argonne National Laboratory\\
Argonne, USA\\
Email: yuri@alcf.anl.gov}
}


%


\maketitle

\begin{abstract}
Quantum Approximate Optimization Algorithm (QAOA) is one of the most promising quantum algorithms for the Noisy Intermediate-Scale Quantum (NISQ) era. Quantifying the performance of QAOA in the near-term regime is of utmost importance. We perform a large-scale numerical study of the approximation ratios attainable by QAOA is the low- to medium-depth regime. To find good QAOA parameters we perform 990 million 10-qubit QAOA circuit evaluations. We find that the approximation ratio increases only marginally as the depth is increased, and the gains are offset by the increasing complexity of optimizing variational parameters. We observe a high variation in approximation ratios attained by QAOA, including high variations within the same class of problem instances. We observe that the difference in approximation ratios between problem instances increases as the similarity between instances decreases. We find that optimal QAOA parameters concentrate for instances in out benchmark, confirming the previous findings for a different class of problems.
\end{abstract}

\begin{IEEEkeywords}
quantum approximate optimization algorithm; max-cut; quantum advantage

\end{IEEEkeywords}

%
\IEEEpeerreviewmaketitle

\section{Introduction}

The Quantum Approximate Optimization Algorithm (QAOA) \cite{farhi2014quantum, farhi2014quantumbounded} is one of the leading candidates for demonstrating quantum advantage, which is the ability to solve a problem faster or find a solution of a higher quality by using a quantum algorithm, compared to classical state-of-the-art alternatives. QAOA requires only shallow circuit depth and can be run on Noisy Intermediate-Scale Quantum (NISQ) devices with limited error correction.
QAOA has been applied to network community detection~\cite{shaydulin2018network,shaydulin2018community}, portfolio optimization~\cite{barkoutsos2019improving} graph maximum cut~\cite{crooks2018performance, zhou2018quantum} and many other problems~\cite{barkoutsos2019improving}. The abundance of work on QAOA applications, as well as the possibility of the speedups over classical state-of-the-art makes quantifying the potential of QAOA all the more urgent and motivates our work.

In this paper, we present a large-scale numerical study of the performance of QAOA on 90 random 10-node Max-Cut instances. For each problem instance we perform extensive (though not exhaustive) search of the variational parameter space, performing 990 million QAOA evaluations in total. We find that the average approximation ratio attained by QAOA on our set of problems is $0.77$ as compared to the ground truth. The maximum approximation ratio we observe is $0.91$. We observe high variation in approximation ratios both between classes of instances and within the same class of instances. We observe that the difference between QAOA approximation ratio grows with the graph edit distance between underlying graphs. We find that QAOA parameters concentrate for the problem instances in our benchmark, indicating that the complexity of QAOA parameter optimization can be addressed by parameter reusing.



\section{The Quantum Approximate Optimization Algorithm (QAOA)}

Consider a Hamiltonian $\Hc$ whose spectrum encodes the solutions of a classical optimization problem. The goal of QAOA is to prepare the state of $\Hc$ corresponding to an optimum of the classical optimization problem, i.e. the highest-energy eigenstate of $\Hc$.

The quantum evolution starts in the uniform superposition over all computational basis states $\ket{+}^{\otimes
n}$. The evolution is performed by applying a series of $p$ alternating operators parameterized by $\vect{\beta},\vect{\gamma}$,

\begin{equation}
    \ket{\psi{(\vect{\beta},\vect{\gamma})}} =  e^{-i\beta_p \HsubM}e^{-i\gamma_p \HsubC}\cdots e^{-i\beta_1 \HsubM}e^{-i\gamma_1 \HsubC}\ket{+}^{\otimes n},
\label{eq:ansatz}
\end{equation}

\noindent where $\Hm=\sum_i\hat{\sigma}_i^x$ is the transverse field mixer Hamiltonian. The role of the classical optimizer is to find variational parameters $\vect{\beta},\vect{\gamma}$ which maximize the expected energy of the cost Hamiltonian,

\begin{equation}
    f(\vect{\beta},\vect{\gamma}) = \bra{\psi{(\vect{\beta},\vect{\gamma})}}\Hc\ket{\psi{(\vect{\beta},\vect{\gamma})}}.
    \label{eq:obj}
\end{equation}

\begin{figure*}[ht]
\subfloat[$p=1$]{\includegraphics[width=\columnwidth]{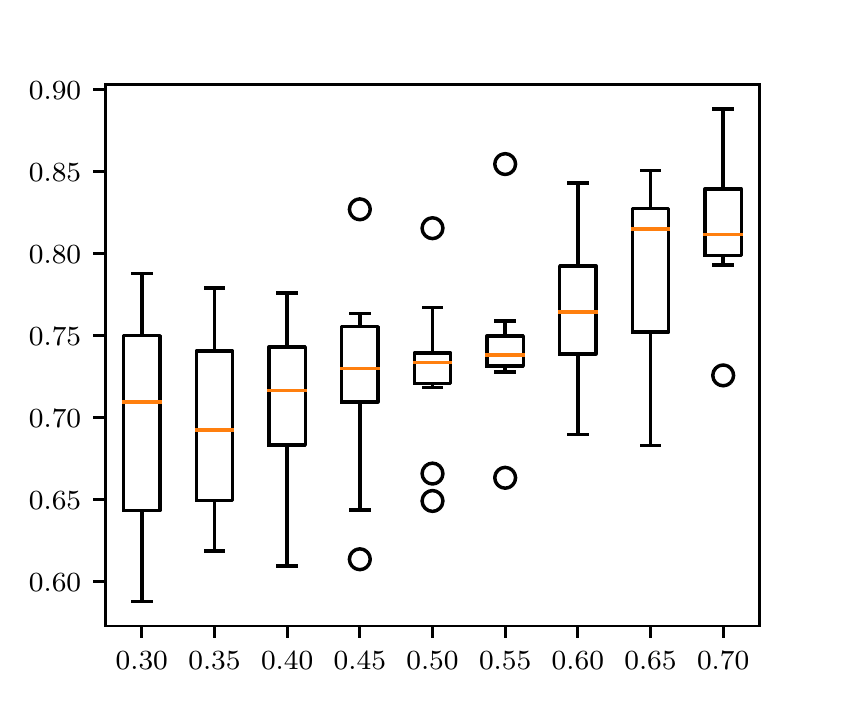}%
\label{fig:approx_d1}}
\hfill
\subfloat[$p=2$]{\includegraphics[width=\columnwidth]{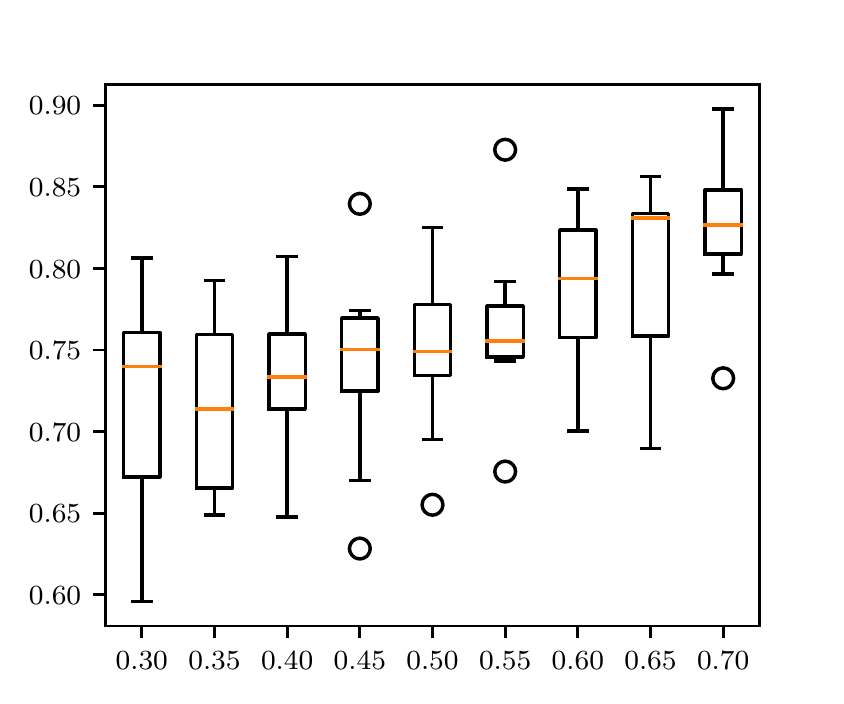}%
\label{fig:approx_d2}}
\caption{Best approximation ratio found by QAOA for different problem classes as a function of edge creation probability $e_p$. (\ref{fig:approx_d1}) presents results for $p=1$ and (\ref{fig:approx_d2}) for $p=2$. (\ref{fig:diff_distance_d1}) presents the absolute difference in QAOA approximation ration as a function of graph edit distance.}
\label{fig:approx_all}
\end{figure*}

The objective function $f$ is periodic with respect to $\vect{\beta}$ and
$\vect{\gamma}$, allowing the parameters to be restricted to $\beta_i\in[0,\pi]$, $\gamma_i\in[0,2\pi]$~\cite{farhi2014quantum}. This gives the optimization domain as $(\vect{\beta},\vect{\gamma})\in
\D=([0,\pi]\times[0,2\pi])^{p}$. For certain problems, the periodicity can be investigated analytically~\cite{PhysRevA.97.022304, zhou2018quantum}, allowing for futher restriction of the optimization domain. However, these results are problem specific.

It has been shown that in the limit $p\rightarrow \infty$, QAOA is capable of finding the true global optimum of the classical combinatorial optimization problem~\cite{farhi2014quantum}. Much less is known about QAOA performance in the low-depth ($1 < p < 10$) depth regime. A recent paper~\cite{crooks2018performance} shows that QAOA can achieve approximation ratios that exceed those of classical Goemans-Williamson~\cite{goemans1995improved} algorithm for Max-Cut. The connection to adiabatic quantum computation can provide an insight into QAOA performance. A number of recent papers explore this connection in depth~\cite{zhou2018quantum, mbeng2019quantum}. Still, there remains a need for a better understanding of QAOA in the low-depth regime, which is motivating our work.

We explore QAOA applied to graph maximum cut (or Max-Cut) problem. Consider a graph $G=(V,E)$, where $V$ is the set of vertices and $E$ is the set of edges. The goal of Max-Cut is to partition of the graph vertices $V$ into two disjoint subsets $V_1$ and $V_2$, $V_1\cup V_2=V$, such that the total number of edges connecting the two subsets is maximized,

\begin{equation} \label{maxcut}
    \max \norm{\{(u,v)\in E \mbox{ s.t. } u\in V_1, v\in V_2 \}} 
\end{equation}

\eqref{maxcut} can be reformulated as \cite{Nannicini2019},

\begin{equation} \label{maxcut2}
    \max_{\vect{s}}\sum_{i,j\in V} w_{ij} s_i s_j + c, \qquad s_k\in \{-1,1\}, \forall k
\end{equation}

\noindent where $w_{ij} = 1$ if $(i,j)\in E$ and $0$ otherwise, and $c$ is a constant. The binary decision variables $s_i$ in \eqref{maxcut2} designate partition membership of the vertices of G after the cut. Finding an exact solution to the Max-Cut problem is known to be NP-hard~\cite{karp1972reducibility}. To solve Max-Cut using QAOA, the cost Hamiltonian is constructed by mapping the binary variables $s_k$ onto eigenvalues of Pauli Z operator $\hat{\sigma}^z$,

\begin{equation}
    \Hc = \sum_{i,j\in V} w_{ij}\hat{\sigma}^z_i\hat{\sigma}^z_j.
\end{equation}

Max-Cut is the most well-studied target problem for QAOA due to the equivalence between Max-Cut and Unconstrained Quadratic Binary Optimization~\cite{maxcut2018}. 

\section{Methods}

We follow Ref.~\cite{zhou2018quantum} in performing extensive searches for QAOA parameters by running many instances of a relatively simple black-box local optimizer. We use derivative-free Bound Optimization BY Quadratic Approximation (BOBYQA)~\cite{powell2009bobyqa} as implemented in the NLopt nonlinear-optimization package~\cite{nlopt}. BOBYQA was shown to perform well for QAOA parameter optimization~\cite{shaydulin2019multistart} as compared to other off-the-shelf derivative-free optimization methods. We set the tolerances on change in the function value to $10^{-3}$ and on the change in optimization parameters to $10^{-2}$. We allow BOBYQA 1 million evaluations for $p=1,2$ and 3 million for $p=4,6,8$. BOBYQA is restarted from a new random point as it converges, with random starting points drawn from a uniform distribution over $\D$. In our experience, with the tolerance levels we use, BOBYQA takes 10-40 iterations to converge, resulting in 20,000-300,000 initial points (exceeding the 10,000 random initial points used in Ref.~\cite{zhou2018quantum}).

Our benchmark consists of 90 Erd{\H{o}}s-R{\'e}nyi random graphs~\cite{erdHos1960evolution} with 10 nodes and edge creation probabilities $e_p$ between $0.3$ and $0.7$ (10 random graphs for each value of $e_p$).  
We use high-performance quantum simulator Qiskit Aer~\cite{Qiskit} for noiseless simulations of QAOA circuits. We use NetworkX~\cite{hagberg2008} for graph manipulations.  We use GNU Parallel for large-scale numerical experiments~\cite{tange_ole_2018_1146014}. All ensemble calculations were performed on Bebop cluster located in Argonne's Laboratory Computing Resource Center (LCRC) and Palmetto cluster at Clemson University. 


\section{Results}

In this section we present the four main findings: 
\begin{enumerate}
    \item Optimization of QAOA parameters becomes challenging for derivative-free black box local optimizers even for relatively low number of steps $p$.
    \item In low-depth regime, average approximation ratios attained by QAOA are limited ($0.77$ for our benchmark), making it challenging to compete with state-of-the-art classical heuristical solvers.
    \item In low-depth regime, approximation ratios exhibit high degree of variability from one problem instance to another even within the same class of instances. The difference in approximation ratio grows with the graph edit distance between problem instances.
    \item We observe strong concentration of optimal QAOA parameters, extending the results presented in Ref.~\cite{brandao2018fixed}
\end{enumerate}

\begin{figure}[htbp]
\centering
\includegraphics[width=\columnwidth]{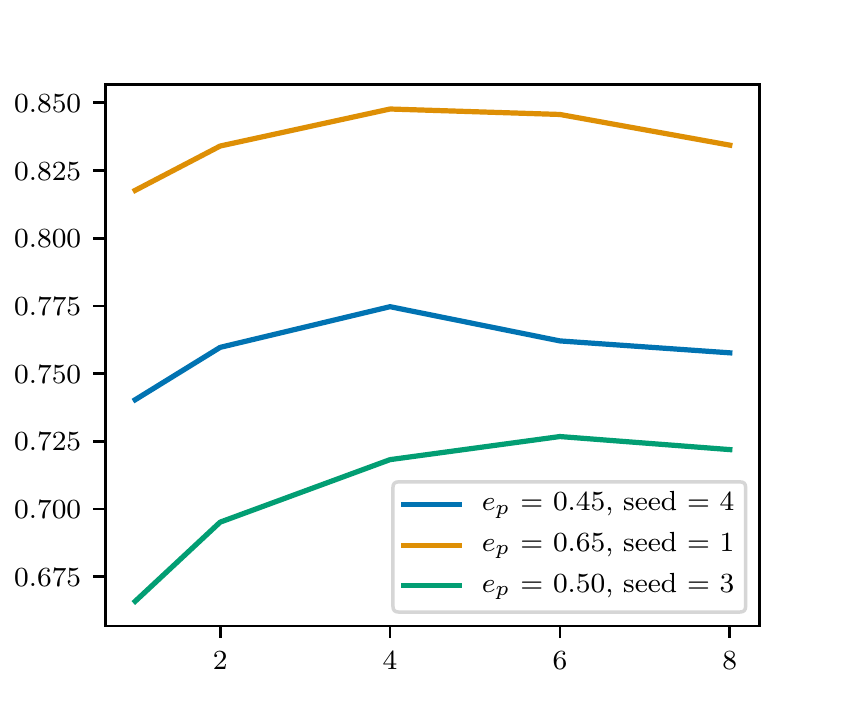}
\caption{Three representative examples of approximation ratio decreasing with the number of QAOA steps due to suboptimality of the parameters $\vect{\beta},\vect{\gamma}$. Three lines correspond to three Erd{\H{o}}s-R{\'e}nyi random graphs with 10 nodes and edge creation probabilities $e_p$. Seed corresponds to NetworkX implementation of Erd{\H{o}}s-R{\'e}nyi random graph generator~\cite{erdosReniyNetworkX}.}
\label{fig:approx_for_depth_example}
\end{figure}

First, we observe that despite considerable budget of evaluations ($1-3$ million evaluations, $\approx 100,000$ initial points) provided to classical optimizer, for $p>4$ we do not obtain optimal variational parameters. As depth of QAOA is increased, the subspace reachable from the initial state is only increased. Therefore the best approximation ratio attained by QAOA with depth $p=k$ is always less or equal than the best approximation ratio for QAOA with depth $p=k+1$.

However, we observe that for $p>4$ the approximation ratios obtained by QAOA are lower than those for $p\leq 4$. We define approximation for problem instance $G$ and a fixed QAOA depth $p$ as the ratio between value of $f$ given best found $\vect{\beta},\vect{\gamma}$ and the ground truth: 

\[
r_{G,p} = \frac{f_{G,p}(\vect{\beta}_{\mbox{opt}},\vect{\gamma}_{\mbox{opt}})}{\mbox{ground truth for }G.}
\]

\begin{figure}[htbp]
\centering
\includegraphics[width=\columnwidth]{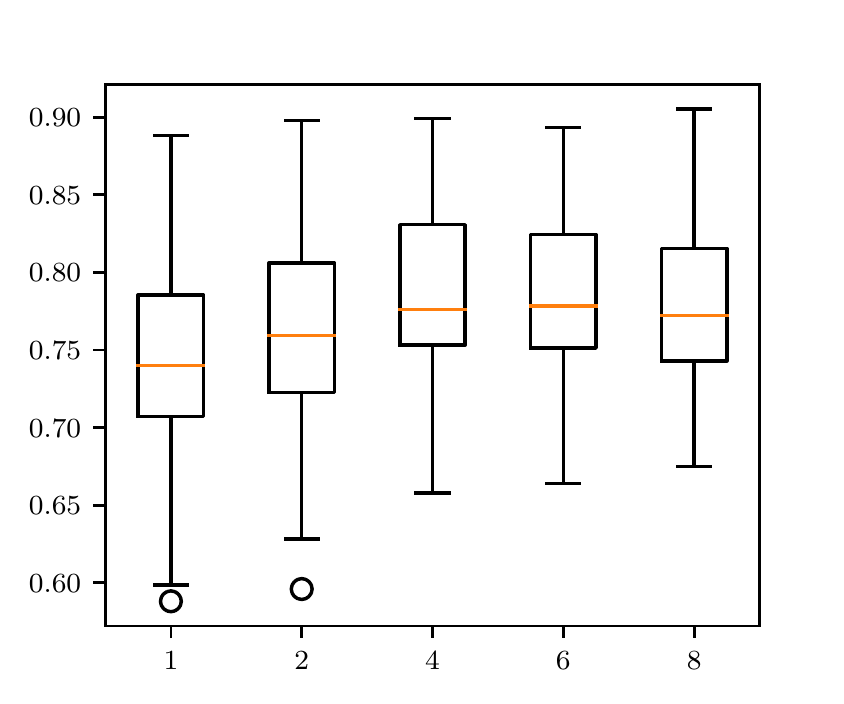}
\caption{Boxplot of approximation ratio as a function of the number of QAOA steps $p$. Median approximation ratio increases for $p\leq 6$ and decreases for $p=8$. High variation of the approximation ratio attained by QAOA can be observed.}
\label{fig:approx_for_depth_boxplot}
\end{figure}

Figure~\ref{fig:approx_for_depth_example} presents this phenomenon on three representative problem instances. Figure~\ref{fig:approx_for_depth_boxplot} presents approximation ratios as a function of depth for the entire dataset. The median approximation ratio increases for $p=1,2,4,6$ and decreases for $p=8$.

Second, we observe that the approximation ratios obtained by QAOA in low-depth regime are limited. Figure~\ref{fig:approx_all} presents the approximation ratios for different classes of problem instances for $p\in\{1,2\}$. We follow~\cite{zhou2018quantum} and examine the optimal parameters obtained for $p\in\{1,2\}$ to confirm that they correspond to the global optimum of $f$. We observe that approximation ratio exhibits high variability within a problem instance class and does not exceed $0.91$.

\begin{figure*}[ht]
\subfloat[$p=1$]{\includegraphics[width=\columnwidth]{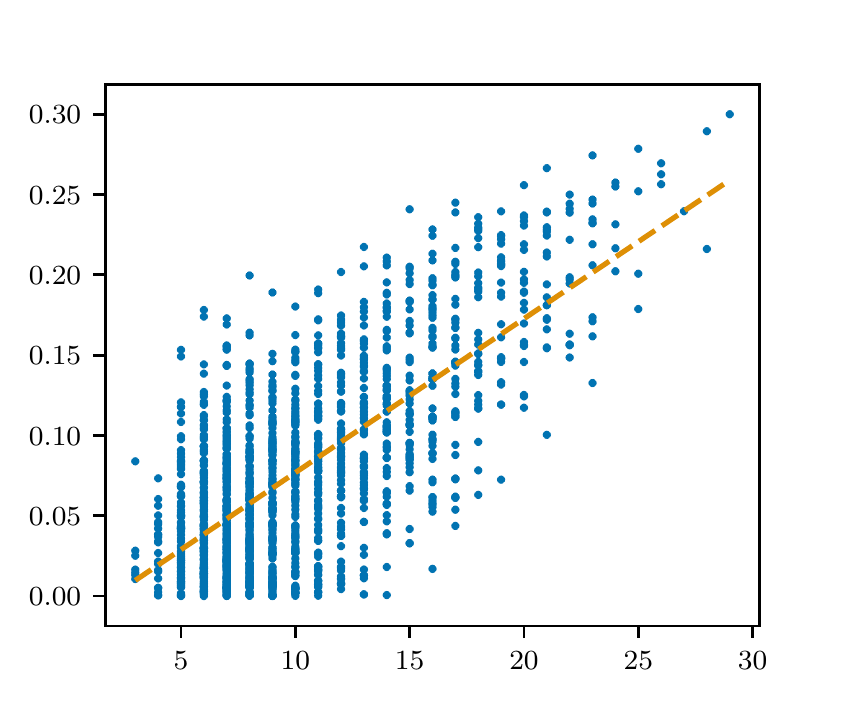}%
\label{fig:diff_distance_d1}}
\hfill
\subfloat[$p=2$]{\includegraphics[width=\columnwidth]{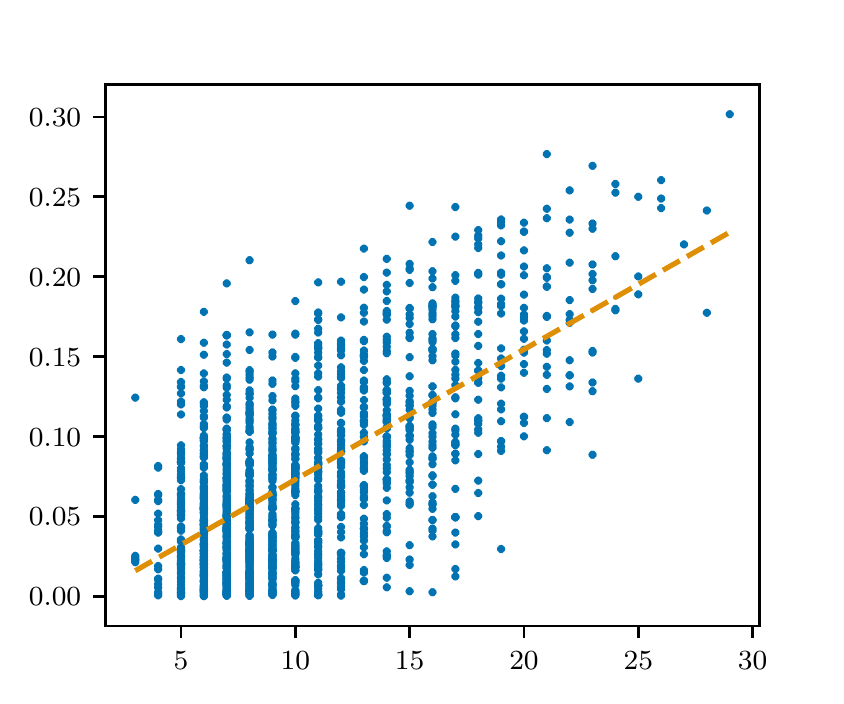}%
\label{fig:diff_distance_d2}}
\caption{The absolute value of the difference in QAOA approximation ratio $d$ as a function of graph edit distance between problem instances. Dashed trend line presents least squares linear fit.}
\label{fig:diff_distance_all}
\end{figure*}

Third, we observe that the difference in approximation ratio obtained by QAOA grows with graph edit distance between problem instances (in other words, QAOA achieves similar approximation ratios for similar problems). Graph edit distance is a graph similarity measure. For graphs $G_1$ and $G_2$, graph edit distance is defined as minimum cost of edit path (a sequence of node and edge operations) transforming $G_1$ into a graph isomorphic to $G_1$. The absolute value of the difference in approximation ratio obtained by QAOA is defined as
\[
d_{G_1, G_2} = |r_{G_1} - r_{G_2}|.
\]

\noindent where $r_{G} = \max_{p\in \{1,2,4,6,8\}}r_{G,p}$. We present $d$ as the function of graph edit distance in Figure~\ref{fig:diff_distance_all}. This observation can have deep implications for machine learning, as it implies that QAOA can be used as a representation that respects similarity between underlying graphs. This warrants further investigation of QAOA as a graph representation tool in machine learning contexts.

\begin{figure*}[htbp]
\subfloat[$p=1$]{\includegraphics[width=\columnwidth]{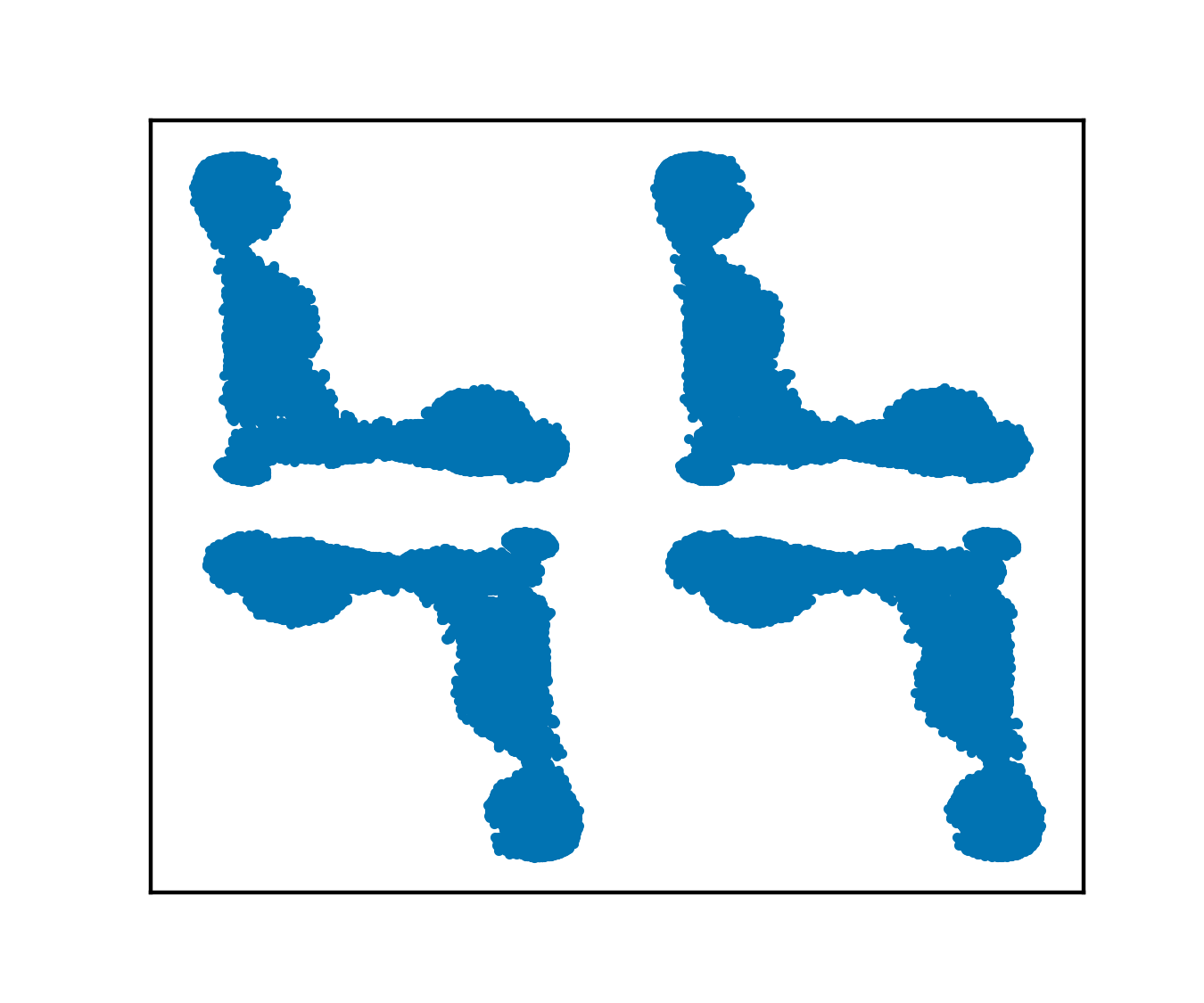}%
\label{fig:best_pts_d1}}
\hfill
\subfloat[$p=2$]{\includegraphics[width=\columnwidth]{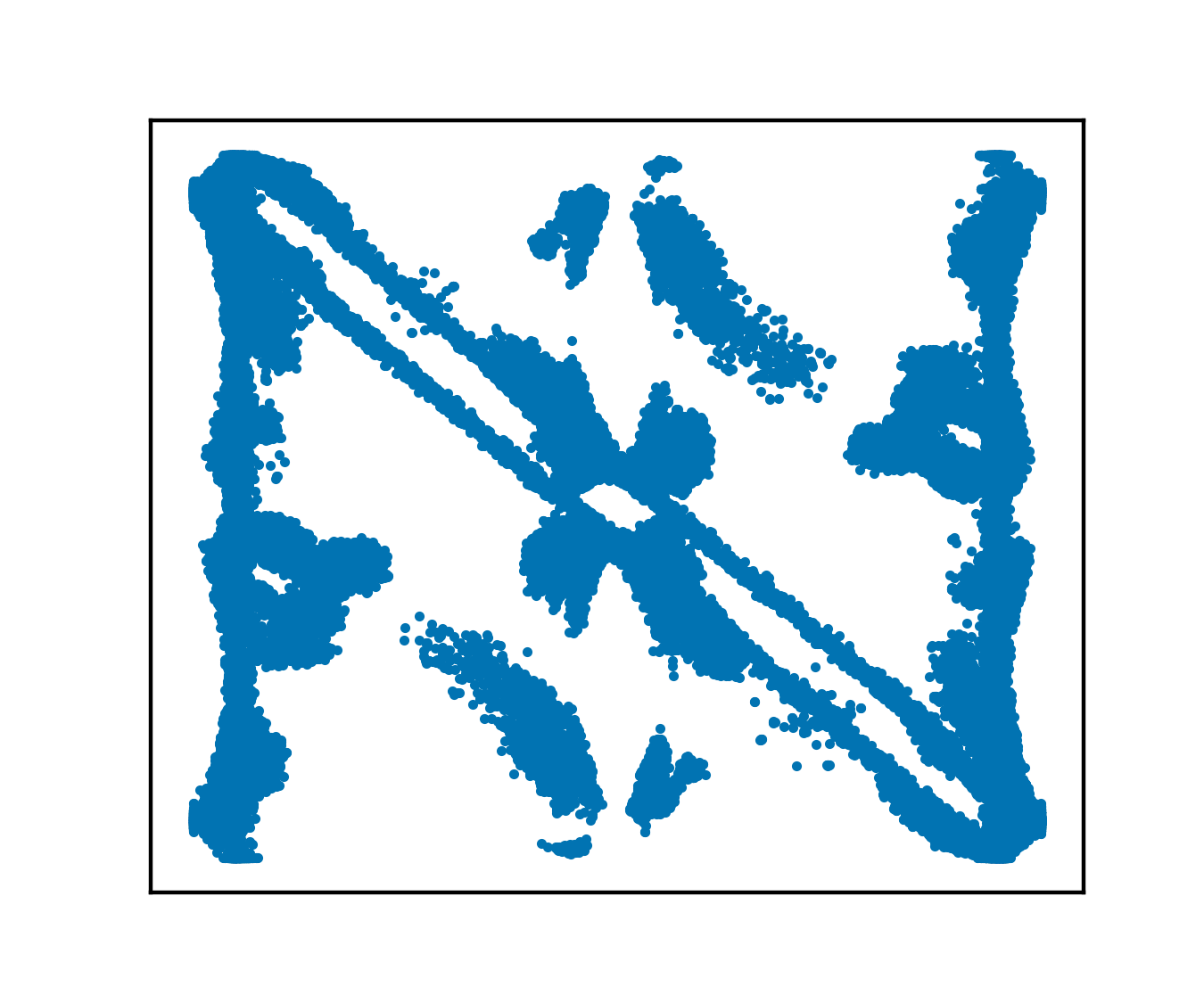}%
\label{fig:best_pts_d2}}
\caption{Parameters $\vect{\beta},\vect{\gamma}$ corresponding to the values of approximation ratio within $1\%$ of the best observed for a given problem instance. For $p=2$ we only plot parameters corresponding to the second QAOA step.}
\label{fig:best_pt_all}
\end{figure*}

Fourth, we observe a strong concentration in optimal parameters. This has been observed previously for Max-Cut on 3-regular graphs~\cite{brandao2018fixed}. Here we extend this observation to Max-Cut on Erd{\H{o}}s-R{\'e}nyi random graphs with unbounded vertex degree. Figure~\ref{fig:best_pt_all} presents the QAOA parameters $\vect{\beta},\vect{\gamma}$ corresponding to the values of approximation ratio within $1\%$ of the best observed for a given problem instance for the entire benchmark (i.e. all values of $e_p$). For $p=2$ (Figure~\ref{fig:best_pts_d2}) we present only the parameters corresponding to the second QAOA step. We observe that the optimal QAOA parameters concentrate around the same values for the problems in our benchmark.

\section{Discussion}

Our results highlight the need for further research into techniques for optimizing QAOA parameters. It is clear that for to achieve good approximation ratios we need to go beyond $p=1,2,4$. At the same time, as the depth increases, the limitations of local optimization methods become evident. The potential of QAOA cannot be realized without advances in variational parameter optimization, including through better understanding of the structure of QAOA objective.

Many recent results provide a path to scaling QAOA to higher $p$. FOURIER heuristic~\cite{zhou2018quantum} is a promising approach, as it is shown to outperform brute force on some classes of Max-Cut problem instances. Combined with other approaches, like multistart methods~\cite{shaydulin2019multistart} and gradient-based backpropagation-inspired approaches~\cite{crooks2018performance}, these methods have the potential to make larger-depth QAOA competitive with classical state-of-the-art heuristics.

Finally, the concentration results presented in Figure~\ref{fig:best_pt_all} suggest that the QAOA training costs can be amortized across a class of problem instances. As optimal parameters concentrate around the same values, it should be possible to fit a model using precomputed optimal parameters for a subset of problem instances and then use that model to efficiently produce optimal QAOA parameters for other problem instances in that class. This approach, originally proposed in~\cite{brandao2018fixed}, has been implemented using reinforcement learning to train a specialized optimizer and Kernel Density Estimation to sample from learned distribution of optimal parameters with no optimization~\cite{saminips,samisc}. This methods can be combined with local optimization heuristics to further improve the performance.

To conclude, there are still numerous challenges to achieving quantum advantage with QAOA. However, we are optimistic that these challenges can be overcome in the near term.

\section*{Acknowledgment}

This work was supported by the Office of Science, U.S. Department of Energy, under Contract DE-AC02-06CH11357. We gratefully acknowledge the computing resources provided on Bebop, a high-performance computing cluster operated by the Laboratory Computing Resource Center at Argonne National Laboratory. Clemson University is acknowledged for generous allotment of compute time on Palmetto cluster.



\bibliographystyle{IEEEtran}
\bibliography{qaoa}
%



\end{document}